\documentclass[10pt,journal,cspaper,compsoc]{IEEEtran}
\usepackage{times}
\usepackage[normalem]{ulem}
\usepackage{array}
\usepackage{multirow}
\usepackage{bbding}
\usepackage{float}
\usepackage{amsmath}
\usepackage{color}
\usepackage{colortbl}
\usepackage{graphicx}
\usepackage{url}
\usepackage{subfigure}
\usepackage{changepage}
\usepackage{epstopdf}
\usepackage[linesnumbered,boxed]{algorithm2e}
\usepackage[justification=centering]{caption}
\definecolor{mygray}{gray}{.88}

\newcommand{\tabincell}[2]{\begin{tabular}{@{}#1@{}}#2\end{tabular}}

\ifCLASSOPTIONcompsoc
\else
\fi
\ifCLASSINFOpdf
\else
\fi
\begin{document}
%
\title{BigDataBench: A Scalable and Unified Big Data and AI Benchmark Suite}
%
%
%
%

\author{Wanling Gao, 
        Jianfeng Zhan, 
        Lei Wang,  
        Chunjie Luo,
        Daoyi Zheng,
        Xu Wen,
        Rui Ren,
        Chen Zheng,
        Xiwen He,
        Hainan Ye,
        Haoning Tang,
        Zheng Cao,
        Shujie Zhang,
        and  Jiahui Dai.
\IEEEcompsocitemizethanks{\IEEEcompsocthanksitem Wanling Gao, Jianfeng Zhan, Lei Wang, Chunjie Luo, Xu Wen, Rui Ren, Chen Zheng and Xiwen He are with State Key Laboratory of Computer Architecture, Institute of Computing Technology, Chinese Academy of Sciences,  University of Chinese Academy of Sciences. E-mail: {gaowanling, zhanjianfeng, wanglei\_2011, luochunjie}@ict.ac.cn.
\IEEEcompsocthanksitem Daoyi Zheng is with Baidu.
Zheng Cao is with Alibaba. Shujie Zhang is with Huawei. Haoning Tang is with Tencent. Hainan Ye and Jiahui Dai are with Beijing Academy of Frontier Sciences and Technology.
\IEEEcompsocthanksitem The corresponding author is Jianfeng Zhan.}
\thanks{}}

\IEEEcompsoctitleabstractindextext{%
\begin{abstract}

Several fundamental changes in technology indicate domain-specific hardware and software co-design is the  only path left. In this context, architecture, system, data management, and machine learning communities
pay greater attention to innovative big data and AI algorithms, architecture, and systems. Unfortunately, complexity, diversity, frequently-changed workloads, and rapid evolution of big data and AI systems raise great challenges. First, the traditional benchmarking methodology that creates a new benchmark or proxy for every possible workload is not scalable, or even impossible for Big Data and AI benchmarking. Second, it is prohibitively expensive to tailor the architecture to characteristics of one or more application or even a domain of applications.

We consider each big data and AI workload as a pipeline of
one or more classes of units of computation performed on different
initial or intermediate data inputs, each class of which we call a data motif. On the basis of our previous work that identifies eight data motifs taking up most of the run time of  a wide variety of big data and AI workloads, we propose a scalable benchmarking methodology that uses the combination of one or more data motifs---to represent diversity of big data and AI workloads.
Following this methodology, we present a unified big data and AI benchmark suite---BigDataBench 4.0, publicly available from~\url{http://prof.ict.ac.cn/BigDataBench}. This unified benchmark suite  sheds new light on domain-specific hardware and software co-design:  tailoring  the system and  architecture to characteristics of the unified eight data motifs other than one or more application case by case. 
Also, for the first time, we comprehensively characterize the CPU pipeline efficiency using the benchmarks of  seven workload types in BigDataBench 4.0 in addition to traditional benchmarks like SPECCPU, PARSEC and HPCC in a hierarchical manner and drill down on five levels, using the Top-Down analysis from an architecture perspective. In addition, we evaluate the micro-architectural performance of AI benchmarks on GPUs.

\end{abstract}

}

\maketitle

\IEEEdisplaynotcompsoctitleabstractindextext

%
\IEEEpeerreviewmaketitle

\section{Introduction}

\IEEEPARstart{S}{everal} fundamental changes in technology, i.e., end of Dennard scaling, ending of Moore's Law, Amdahl's Law, and its implications for ending `easy' multi-core era, indicate domain-specific hardware and software co-design is the  only path left~\cite{2018_turing, gao2018motif}. Among many domains, Big Data and AI are the brightest star in the sky, and hence
architecture, system, data management, and machine learning communities
pay greater attention to innovative big data and AI algorithms, architecture, and systems~\cite{barroso2009datacenter, ferdman2011clearing, jouppi2017datacenter, abadi2016tensorflow, jia2014caffe, bigbench, wang2010transformer,wang2014bigdatabench, jia2016auto, luo2017cosine, zhou2017pdeep}. 
Unfortunately, complexity, diversity, frequently-changed workloads---so called workload churns~\cite{barroso2009datacenter}, and rapid evolution of big data and AI systems raise great challenges in benchmarking 
and domain-specific hardware and software co-design.

\begin{table*}[htbp]
\scriptsize
\centering
\caption{The Summary of Different Big Data
Benchmarks.}\label{comparition_table}
\center
\begin{tabular}{|p{0.9in}|p{0.93in}|p{0.85in}|p{0.45in}|p{0.35in}|p{0.4in}|p{1.1in}|p{0.35in}|}
\hline
&\tabincell{c}{Benchmarking Target} &\tabincell{c}{Methodology} &\tabincell{c}{Application\\domains}
&\tabincell{c}{Workload\\types} &\tabincell{c}{Workloads}
&\tabincell{c}{Scalable data sets abs-\\tracting from real data}
&\tabincell{c}{Software\\Stacks}\\
\hline
\tabincell{l}{BigDataBench 4.0}& \tabincell{l}{Big data and AI sys-\\tems and architecture} & Motif-based & 5 & 7\footnotemark[1] & \tabincell{l}{47} & \tabincell{l}{13 real data sets;\\6 scalable data sets} & \tabincell{l}{17}\\
\hline
\tabincell{l}{BigDataBench 2.0\\~\cite{wang2014bigdatabench}} & \tabincell{l}{Big data systems and\\ architecture} & Popularity & 3 & 3 & 19 & \tabincell{l}{6 real data sets;\\ 6 scalable data sets} & \tabincell{l}{10}\\
\hline
\tabincell{l}{BigBench 2.0~\cite{rabl2015vision}} & Big data systems & Application model & 1 & 5 & Proposal & Proposal & Proposal\\
\hline
\tabincell{l}{BigBench 1.0~\cite{bigbench}} & Big data analytics & Application model & 1 & 1 & 10 & 3 data generators & 3\\
\hline
\tabincell{l}{CloudSuite 3.0~\cite{ferdman2011clearing}} & Cloud services & Popularity & N/A & 4 & 8 & 3 data generators & 3\\
\hline
\tabincell{l}{HiBench 6.0~\cite{huang2010hibench}} & Big data systems & Popularity & N/A & 6 & 19 & Random generate or with specific distribution & 5\\
\hline
CALDA~\cite{pavlo2009comparison}& MapReduce system and parallel DBMSs  & Popularity &N/A& 1 & 5 &N/A&\tabincell{l}{3}\\
\hline
YCSB~\cite{ycsb}& Cloud serving systems & Performance model &N/A& 1 & 6 &N/A& 4 \\
\hline
LinkBench~\cite{armstrong2013linkbench}& Database systems & Application model &N/A& 1 & 10 & 1 data generator& 2\\
\hline
AMP Benchmarks~\cite{amplabbenchmark}& Data analytic systems & Popularity &N/A& 1 & 4 &N/A&\tabincell{l}{5}\\
\hline
Fathom~\cite{adolf2016fathom} & AI systems & Popularity & N/A & 1 & 8 & N/A & 1\\
\hline
DeepBench~\cite{deepbench} & AI systems & Popularity & N/A & 1 & 4 & N/A & 1 \\
\hline
BenchNN~\cite{chen2012benchnn} & AI systems & Popularity & N/A & 1 & 5 & N/A & 1 \\
\hline
DNNMark~\cite{dong2017dnnmark} & AI systems & Popularity & N/A & 1 & 8 & N/A & 1\\
\hline
DAWNBench ~\cite{coleman2017dawnbench} & AI systems & Popularity & N/A & 1 & 2 & N/A & 2\\
\hline
\end{tabular}

\footnotemark[1]{The seven workload types are online service, offline analytics, graph analytics, artificial intelligence (AI), data warehouse, NoSQL, and streaming.}

\end{table*}

The traditional  benchmark methodology that creates a new benchmark or proxy for every possible workload is prohibitively costly and hence not scalable~\footnote{The meaning of scalable differs from scaleable. As one
of four properties of domain-specific benchmarks defined by Jim Gray~\cite{gray1992benchmark}, the latter refers to scaling the benchmark up to larger systems}, or even impossible for Big Data and AI benchmarking. 
First, there are many classes of big data and AI applications. Even
for Internet services, there are several important
application domains, e.g., search engines, social networks,
and e-commerce. The value of big data and AI
also drives the emergence of innovative application domains. Meanwhile, data (sizes, types, sources, and patterns) have a great impact on workload behaviors and performance significantly~\cite{gao2018motif, xie2018cvr}, so comprehensive and representative real-world data sets should be included.

Second,
at an earlier stage, it is usually difficult to justify porting a full-scale end-to-end Big data or AI application to a new computer system or  architecture simply to obtain a benchmark number~\cite{bailey1991parallel}; while at a later stage, kernels alone are insufficient
to completely assess the performance potential of a new system or architecture  on real-world data sets and applications~\cite{bailey1991parallel}.  Meanwhile, the benchmarks should be consistent across different communities for the co-design of software and hardware. 

Third, the correctness of results and performance figures
must be easily verifiable~\cite{bailey1991parallel}. To some extent, too complex  workloads, i.e., full-scale end-to-end Big Data or AI applications  raise difficulties in  reproducibility and interpretability of performance data~\cite{gao2018motif}.

As modern big data and AI workloads are not only
diverse, but also fast changing and expanding, it also raises great challenges in domain-specific hardware and software co-design. Even the agile hardware development methodology and tools are adopted~\cite{2018_turing}, it is prohibitively expensive to tailor the architecture to characteristics of one or more application or even a domain of applications, and hence building domain-specific hardware and software systems case by case should be avoided.

This paper presents our joint research efforts on a scalable and unified Big Data and AI benchmarking suite with several industrial partners.
 On the basis of our previous work~\cite{gao2018motif} that identifies eight data motifs--- taking up most of the run time among  a wide variety of big data and AI workloads, we propose a scalable benchmarking methodology that uses the combination of one or more data motifs---including \emph{Matrix, Sampling, Transform, Graph, Logic, Set, Sort and Statistic computation} to represent diversity of big data and AI workloads. Our benchmark suite includes  micro benchmarks, each of which  is a single data motif, the component benchmarks, each of which consists of the combination of one or more data motifs with different weights in terms of runtime, and end-to-end application benchmarks, which are combinations of component benchmarks. 

Following this methodology, we present a unified big data and AI benchmark suite---BigDataBench 4.0, publicly available from~\url{http://prof.ict.ac.cn/BigDataBench}.
BigDataBench 4.0 provides 13 representative real-world data sets and 47 big data and AI benchmarks of seven workload types: online service, offline analytics, graph analytics, AI, data warehouse, NoSQL, and streaming. Also, for each workload type, we provide diverse implementations using state-of-the-art and state-of-the-practise software stacks.
Data varieties are considered with the whole spectrum of data
types including structured, semi-structured, and unstructured
data. 
Using real data sets as the seed, the data generators \cite{ming2014bdgs} are provided to generate the data with a specific scale.

On a typical state-of-practice processor: Intel Xeon E5-2620 V3, we comprehensively characterize the benchmarks of seven  workload types in \emph{BigDataBench}  in addition to SPECCPU, PARSEC, and HPCC using the Top-Down method ~\cite{yasin2014top}. We classify an issued micro operation (uops) into \emph{retiring, bad speculation, frontend bound and backend bound}, among which, only \emph{retiring} represents useful work.
In order to explore AI workloads' characteristics thoroughly, we run them on both CPUs and GPUs to evaluate their micro-architectural performance.

We have the following observations.
First, the average ILP (instruction-level parallelism) and MLP (memory-level parallelism) of the AI benchmarks
are almost 1.5 times higher than that of Big Data.
 With respect to the traditional benchmarks, i.e., SPECCPU, PARSEC, and HPCC, the average ILP of AI is lower, and the AI framework needs more optimizations like instruction mix balance and memory access locality.

Second, in terms of uppermost-level breakdown, AI reflect similar pipeline behaviors with the traditional benchmarks.
However, to explore deeply, their bottlenecks that incur the frontend and backend stalls are different, which means AI benchmarks have distinct computation patterns comparing to the traditional benchmarks.
Corroborating the observations in the previous work~\cite{ferdman2011clearing,kanev2015profiling,jia2017understanding}, the frontend bound of Big Data is more severe than that of the traditional benchmarks.
 However, we notice that the frontend bound varies across  different workload types.

Third, for Big Data and AI, they have more CISC instructions that cannot be decoded by default decoders, almost 10 times larger than that of the traditional benchmarks. So they suffer from more penalties because of switching to a special unit.
Fourth, corroborating the previous work~\cite{jia2017understanding}, the first bottleneck is backend bound for Big Data and AI. However, different from the previous work~\cite{jia2017understanding}, we observe that the  data movement delay among memory hierarchies is the main reason for backend bound, especially the latency delay from DRAM memory.
Fifth, the utilizations of GPU resources vary when running different AI benchmarks. The stalls because of the data movements limit their performance on GPUs. In addition, the iteration number has little impact on architectural behaviors of AI.

In summary, our contributions are three-fold as follows.

\hangafter 1
\hangindent 2.5em
1)  We propose a data motif-based scalable benchmarking methodology.

\hangafter 1
\hangindent 2.5em
2)  We present a unified big data and AI benchmark suite---BigDataBench 4.0.

\hangafter 1
\hangindent 2.5em
3) We thoroughly perform workload characterizations of big data and AI benchmarks on CPUs and GPUs, respectively.
\vspace{0.1ex}

The rest of this paper is organized as follows. In Section 2, we present the related work and background. Section 3 summarizes our benchmarking methodology. Section 4 presents our unified Big Data and AI benchmark suite----\emph{BigDataBench 4.0}.
Section 5 illustrates the experiment configurations. In Section 6, we present the characterization results.
Finally, we draw the conclusion in Section 7.

\begin{figure*}[tb]
\centering
\includegraphics[scale=0.25]{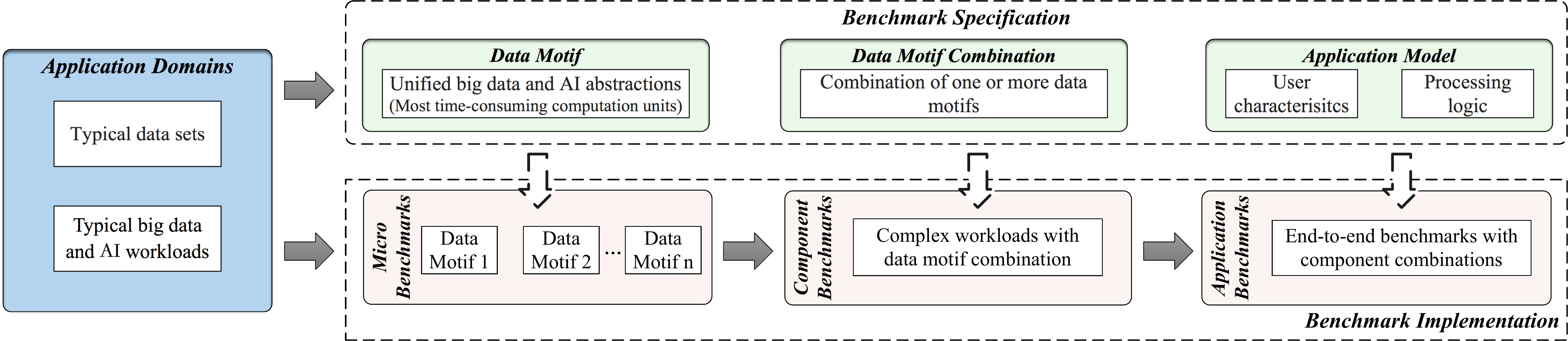}
\caption{BigDataBench 4.0 Methodology.} 
\label{BigDataBenchM}
\end{figure*}

\section{Related Work}

Identifying units of computation in corresponding domains is an important step towards understanding various workloads~\cite{colella2004defining,asanovic2006landscape,gao2018motif}. TPC-C~\cite{chen2014tpc} benchmark is based on the units of computation in the OLTP domain. HPCC~\cite{luszczek2006hpc} benchmark abstracts seven basic operations in high performance computing.
Following the `pencil and paper' specification, the NAS parallel benchmark~\cite{bailey1991parallel} consists of five `parallel kernel' benchmarks and three `simulated application' benchmarks, and together they mimic the computation and data movement
characteristics of large-scale computational fluid
dynamics applications. Our previous work~\cite{gao2018motif} identifies eight data motifs among a wide variety of  big data and AI workloads, including \emph{Matrix, Sampling, Transform, Graph, Logic, Set, Sort} and \emph{Statistic} computations. National Research Council identifies seven major tasks in massive data analysis~\cite{council2013frontiers}, which are macroscopical definition of problems from mathematic perspective.
Fox et al~\cite{fox2015big} build a set of Big Data application characteristics with 50 features, which they call facets and divide them into 4 views. As a machine learning framework, TensorFlow~\cite{abadi2016tensorflow} adopts a dataflow-based programming abstraction, using individual mathematical operators as nodes in the dataflow graph.
Cambricon~\cite{liu2016cambricon} is an instruction set architecture for neural networks, which is abstracted from instruction level. 

Big Data and AI attract great attention, appealing many research efforts on big data and AI benchmarking, as illustrated in Table~\ref{comparition_table}.
Our previous work---BigDataBench 2.0~\cite{wang2014bigdatabench} abstracts three application domains and provides nineteen workloads covering offline analytics, online services and data warehouse, which targets big data systems and architecture.
BigBench 1.0 \cite{bigbench} targets a product retailer business model based on TPC-DS \cite{tpcds} and targets big data analytics workloads. BigBench 2.0~\cite{rabl2015vision} is a proposal which still focuses on retail business model and adds four workload types of streaming, key-value processing, graph processing, and a multimedia data type.
CloudSuite 3.0~\cite{ferdman2011clearing} is a benchmark suite for cloud service,  and choose workloads according to popularity, totally including four workload types and eight workloads. It evaluated the server inefficiencies from the frontend and backend, however, the analysis did not drill down on the deeper levels.
HiBench 6.0 \cite{huang2010hibench} also chooses workloads according to popularity, containing six workload types and nineteen workloads, including micro benchmarks, machine learning, sql, graph, websearch and streaming categories.
YCSB \cite{ycsb} released by Yahoo! is a benchmark for data storage systems and only includes online service workloads, i.e. Cloud OLTP. The workloads are mixes of read/write operations to cover a wide performance space.
CALDA \cite{pavlo2009comparison} is a benchmarking effort targeting  MapReduce systems and parallel DBMSes. Its workloads are from the original MapReduce paper~\cite{dean2008mapreduce} and add four complex analytical tasks.
LinkBench \cite{armstrong2013linkbench} is a synthetic benchmark for database systems which models the data scheme and workload patterns according to Facebook.
AMP benchmark~\cite{amplabbenchmark} is a big data benchmark proposed by UC Berkeley, which focuses on real-time analytic applications. The workloads are from CALDA benchmark.

A series of AI benchmarks are proposed as follows. 
Fathom~\cite{adolf2016fathom} provides eight deep learning workloads implemented with TensorFlow.
DeepBench~\cite{deepbench} consists of four operations involved in training deep neural networks, including three basic operations and recurrent layer types.
BenchNN~\cite{chen2012benchnn} develops and evaluates software neural network implementations of 5 (out of 12) high-performance applications from the PARSEC Benchmark Suite.
DNNMark~\cite{dong2017dnnmark} is a GPU benchmark suite that consists of a collection of deep neural network primitives.
Tonic Suite~\cite{hauswald2015djinn} presents seven neural network workloads that use the DjiNN service.
DAWNBench~\cite{coleman2017dawnbench} is a benchmark and competition focusing on end-to-end training time to achieve a state-of-the-art accuracy level, as well as inference time with that accuracy. It focuses on two tasks including image classification on CIFAR10 and ImageNet, and question answering on SQuAD.
SLAB (Scalable Linear Algebra Benchmarking)~\cite{thomascomparative} presents a suite of LA-specific tests based on the analysis of data access and communication patterns of LA workloads.

\section{Data motif-based scalable benchmarking Methodology}\label{methodology}

In this section, we introduce our data motif-based scalable benchmarking methodology.

We consider each big data and AI workload as a pipeline of
one or more classes of units of computation performed on different
initial or intermediate data inputs~\cite{gao2018motif}. Each class of unit of computation captures the
common requirements while being specified only algorithmically in a `paper-and-pencil' approach~\cite{bailey1991parallel} and reasonably divorced from
individual implementations~\cite{asanovic2006landscape}, and hence we call it a data motif~\cite{gao2018motif}. Significantly different from the traditional
kernels~\cite{bailey1991parallel}, a data motif's behaviors are affected by its data sizes, patterns,
types, and sources, reflecting not only computation patterns, memory access patterns, but also disk and
network I/O patterns~\cite{gao2018motif}.

\subsection{Background of Eight Data Motifs}\label{motifs}


After profiling forty big data and AI workloads with a broad spectrum, our previous work identifies eight unified data motifs among big data and AI workloads,including \emph{Matrix, Sampling, Transform, Graph, Logic, Set, Sort} and \emph{Statistic} computations.
Among them, matrix computation involves vector-vector, vector-matrix and matrix-matrix computations. Sampling is a method to select a subset of original data from within a statistical population.
Transform computation indicates the conversion from the original domain to another domain, such as FFT. Graph computation uses nodes representing entities and edges representing dependencies.
Logic computation performs bit manipulation.
Set computation means the operations on one or more collections of data. Please note that primitive operators in relation algebra~\cite{codd1970relational} are also classified into set computation in our motif taxonomy.
Sort and statistic computation are fundamental units of computation in big data and AI. For online services, get, put, post, and delete are identified as basic and abstract operations in the previous work~\cite{guinard2010resource}, so we use them directly to construct online service benchmarks and don't include those four in our motif set.

\subsection{Benchmarking Methodology}

Fig. \ref{BigDataBenchM} summarizes our data motif-based scalable benchmarking methodology for BigDataBench 4.0, separating the specification from implementation.
First, through investigating typical application domains using some widely acceptable metrics, e.g. page views for internet service, we thoroughly analyze these domains in terms of processing logic and data pipeline.
Second, we choose representative workloads from these domains. After profiling these workloads, we analyze their computation dependency graph and run time breakdown, and find the hotspot functions. Combing with algorithmic analysis, we decompose the workloads and summarize the frequently-appearing and time-consuming units of computation within these workloads as data motifs~\cite{gao2018motif}.
Finally, circling around the data motifs identified from these application domains, we then define the specifications of micro, component, and end-to-end application benchmarks, as the guidelines for benchmark implementation. 
The specifications of micro, component, and application benchmarks are as follows.

\textbf{Micro Benchmark Specification}
As illustrated in Subsection \ref{motifs},
data motifs are fundamental concepts and unified units of computation among a majority of big data and AI workloads.
We design a suite of micro benchmarks, each of which is a single data motif, widely used in investigated application domains, as listed in Table~\ref{BigDataBench_micro}.

\begin{figure}[!t]
\centering
\includegraphics*[scale=0.4, angle=90]{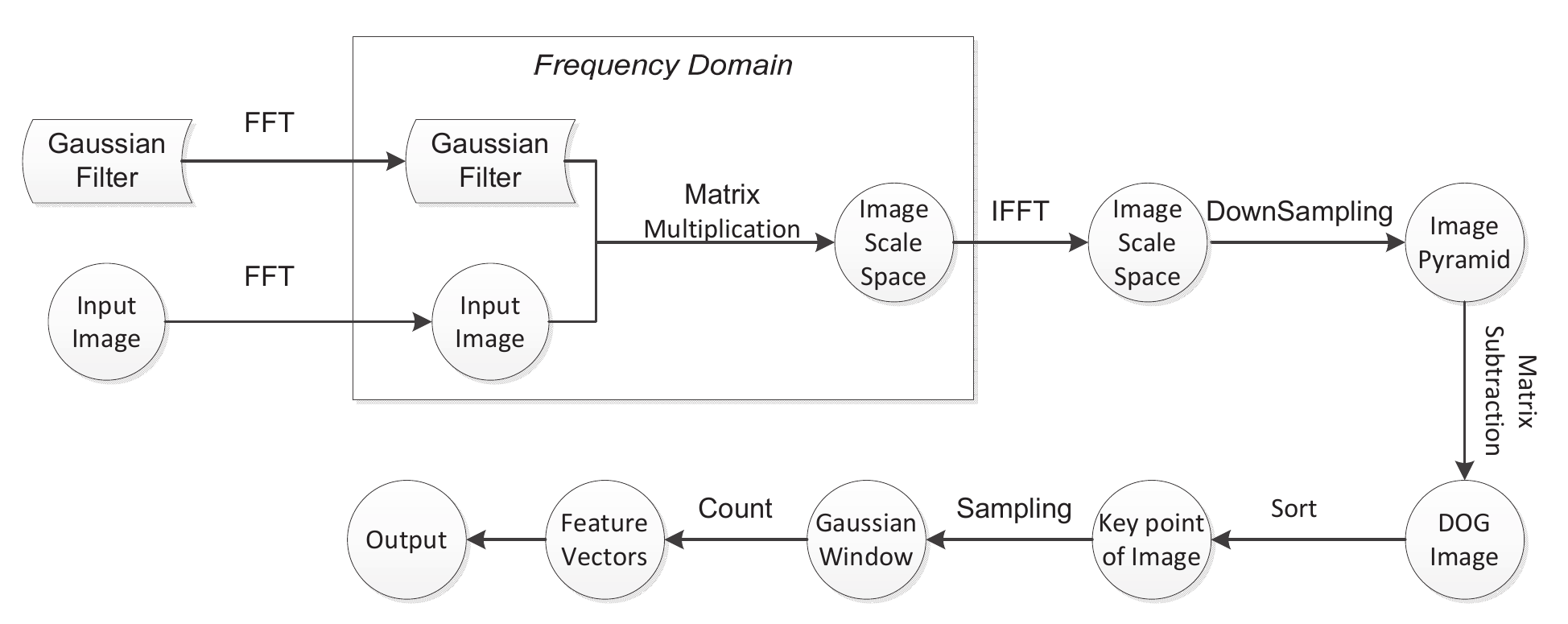}
\caption{The DAG-like Structure of SIFT Algorithm. SIFT Benchmark is a combination of several motifs: transform (FFT, IFFT), sampling (downsampling), matrix (matrix multiplication/subtraction), sort (sort), statistics (count).} 
\label{SIFT}
\end{figure}

\textbf{Component Benchmark Specification}
Considering the benchmarking scalability, we use the motif combinations to compose original complex workloads with a DAG-like structure considering the data pipeline. 
The DAG-like structure is to use a node representing original or intermediate data set being processed, and an edge representing a data motif.
Table \ref{BigDataBench_component} lists the component benchmarks. For example, SIFT~\cite{lowe2004distinctive} is a combination of five data motifs, including matrix, sampling, transform, sort and statistic computations, Fig. \ref{SIFT} presents its DAG-like structure, which specifies how data set or intermediate data set are operated by different motifs.


\begin{figure}[tb]
\centering
\includegraphics[scale=0.39]{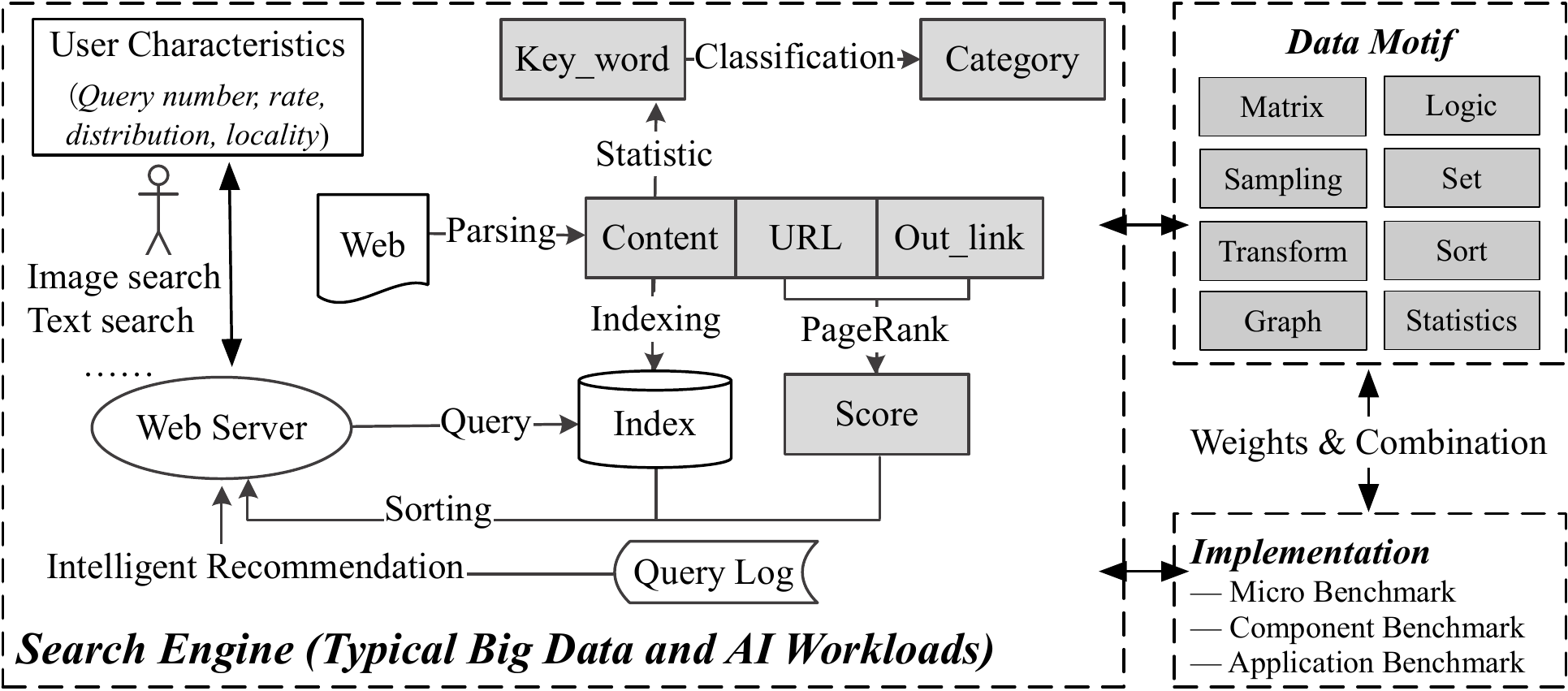}
\caption{Benchmark Specification and Implementation: Search Engine Example.} 
\label{se-spec}
\end{figure}

\begin{table*}[htbp]
\scriptsize
\caption{The Summary of Micro Benchmarks in BigDataBench 4.0.}
\renewcommand\arraystretch{1.4}
\label{BigDataBench_micro}
\center 
\begin{tabular}{|p{0.9in}|p{0.65in}|p{0.9in}|p{0.7in}|p{1.1in}|p{1.5in}|}
\hline
\textbf{Micro Benchmark} &  \textbf{Involved Motif} & \textbf{Application Domain} & \textbf{Workload Type} & \textbf{Data Set} & \textbf{Software Stack}\\
\hline
Sort & Sort & \multirow{5}{*}{\tabincell{l}{SE, SN, EC, MP, BI}} & Offline analytics & Wikipedia entries & Hadoop, Spark, Flink, MPI\\
\cline{1-2}\cline{4-6}
\multirow{2}{*}{Grep} & \multirow{2}{*}{Set} &  & Offline analytics & Wikipedia entries & Hadoop, Spark, Flink, MPI\\
\cline{4-6}
& & & Streaming & Random Generate & Spark streaming\\
\cline{1-2}\cline{4-6}
WordCount & Basic statistics &  & Offline analytics & Wikipedia entries & Hadoop, Spark, Flink, MPI\\
\cline{1-2}\cline{4-6}
MD5 & Logic &  &  Offline analytics & Wikipedia entries & Hadoop, Spark, MPI\\
\hline
Connected Component & Graph & SN & Graph analytics  & Facebook social network & Hadoop, Spark, Flink, GraphLab, MPI\\
\hline
RandSample & Sampling & SE, MP, BI &  Offline analytics & Wikipedia entries & Hadoop, Spark, MPI\\
\hline
FFT & Transform & MP & Offline analytics & Two-dimensional	matrix & Hadoop, Spark, MPI\\
\hline
\tabincell{l}{Matrix Multiply} & Matrix & \tabincell{l}{SE, SN, EC, MP, BI} & Offline analytics & Two-dimensional matrix & Hadoop, Spark, MPI\\
\hline
Read / Write / Scan & Set & SE, SN, EC & NoSQL & ProfSearch resumes & HBase, MongoDB\\
\hline
Convolution & Transform & SN, EC, MP, BI & AI & Cifar, ImageNet & TensorFlow, Caffe, PyTorch\\
\hline
Fully Connected & Matrix & SN, EC, MP, BI & AI & Cifar, ImageNet & TensorFlow, Caffe, PyTorch\\
\hline
Relu & Logic & SN, EC, MP, BI & AI & Cifar, ImageNet & TensorFlow, Caffe, PyTorch\\
\hline
Sigmoid / Tanh & Matrix & SN, EC, MP, BI & AI & Cifar, ImageNet & TensorFlow, Caffe, PyTorch\\
\hline
MaxPooling & Sampling & SN, EC, MP, BI & AI & Cifar, ImageNet & TensorFlow, Caffe, PyTorch\\
\hline
AvgPooling & Sampling & SN, EC, MP, BI & AI & Cifar, ImageNet & TensorFlow, Caffe, PyTorch\\
\hline
CosineNorm~\cite{luo2017cosine} & Basic Statistics & SN, EC, MP, BI & AI & Cifar, ImageNet & TensorFlow, Caffe, PyTorch\\
\hline
BatchNorm~\cite{ioffe2015batch} & Basic Statistics & SN, EC, MP, BI & AI & Cifar, ImageNet & TensorFlow, Caffe, PyTorch\\
\hline
Dropout~\cite{srivastava2014dropout} & Sampling & SN, EC, MP, BI & AI & Cifar, ImageNet & TensorFlow, Caffe, PyTorch\\
\hline

\end{tabular}
\end{table*}

\textbf{Application Benchmark Specification}
To model an application domain, we define an end-to-end application benchmark specification considering user characteristics and processing logic, based on the real process of an application domain. 
We abstract the primary processes of an application domain, and then further propose portable and usable end-to-end benchmarks. In benchmarking, we also consider user characteristics. For example, for online service, we generate queries considering query number, rate, distribution and locality to reflect the user characteristics.

Due to the space limitation, we take search engine as an example and illustrate our methodology to construct benchmarks. As shown in Fig.~\ref{se-spec}, we first abstract a search engine application model, including the online search server (e.g. image search, text search), and offline analytics (e.g. indexing, classification, recommendation). From the algorithm and profiling levels, we identify the involved data motifs mainly used in search engine. Then we define benchmark specification from three levels: 1) choosing the single data motif as micro benchmark, such as sort, statistics; 2) choosing data motif combinations with different weights as primary component benchmarks in search engine, such as pagerank, index, search server; 3) combing component benchmarks to build a search engine with processing logic as application benchmark.

\subsection{Why a Scalable Benchmarking Methodology}

Traditional benchmarking methodology provides a case-by-case solution and creates a new benchmark for each workload. However, it is costly and even impossible due to the complexity and diversity of big data and AI applications. Moreover, the emergence of innovative applications aggravates this issue and brings great difficulties and development costs in order to keep in pace.

NAS benchmark~\cite{bailey1991parallel} adopts a ``paper and pencil" specification, which specifies a set of problems only algorithmically and provides kernel-based benchmarks. However, kernel-based methodology is insufficient for big data and AI benchmarking, considering the data varieties.

Our benchmarking methodology is a significant departure from the traditional benchmark methodology. First, for the sake of conciseness, representativeness, and benchmarking cost, our methodology captures the common classes of units of computation, easily combine a new workload, and hence it is scalable. Second, at an earlier stage, it is easy to port micro benchmarks to a new computer system or architecture, while at a later stage, component benchmarks and application benchmarks are sufficient for completely performance evaluations. Third, the evaluation results of data motif-based benchmarks are easily reproducible and verifiable, because of the interpretability of data motif behaviors.

\section{Unified Big Data and AI benchmark suite}

In this section, first, we discuss why we propose a unified benchmark suite, and then we summarize our benchmark decisions in  BigDataBench 4.0.

\subsection{Why a Unified Benchmark Suite}

There are three reasons for why we need a unified benchmark suite for both Big Data and AI. 
First, being specified  algorithmically in a `paper-and-pencil' approach~\cite{bailey1991parallel}, we can state the common requirements of both Big Data and AI. 
Second, the unified benchmark suite sheds new light on  domain-specific hardware and software co-design in terms of tailoring  the system and architecture to characteristics of data motifs other than one or more application case by case.
Third, the unified benchmark suite helps performing an apple-to-apple comparison on different system and architecture implementations.

\subsection{Benchmark Decisions}



On the basis of the benchmarking methodology, we make benchmark decisions and build BigDataBench 4.0. 
As there are many emerging big data and AI applications, we take an incremental and iterative approach.
We choose five important and emerging application domains according to occupancy and growing rate. Search engine, social network, e-commerce from internet service, occupy 80\% page views and daily visitors~\cite{alexatop500}. Multimedia processing and bioinformatics are emerging big data domains~\cite{multimediagrowth,bioinformatics}.
Then we build domain-specific benchmarks considering workload, data, and state-of-the-art techniques. 

\subsubsection{Workloads Diversity}

After investigating fundamental components in application domains, we provide a suite of micro benchmarks and component benchmarks.
Table \ref{BigDataBench_micro} and Table \ref{BigDataBench_component} present the micro and component benchmarks of BigDataBench 4.0 respectively, from perspectives of workloads, involved data motifs, application domains, workload types, data sets and software stacks.
Note that we use SE, SN, EC, MP and BI for short to represent search engine, social network, e-commerce, multimedia processing and bioinformatics domains, respectively.
Totally, we provide 47 big data and AI benchmarks, each of which has diverse implementations. Because of the page limitation, we do not report the application benchmarks.


\begin{table*}[htbp]
\scriptsize
\caption{The Summary of Component Benchmarks in BigDataBench 4.0.}
\renewcommand\arraystretch{1.4}
\label{BigDataBench_component}
\center
\begin{tabular}{|p{0.8in}|p{1.38in}|p{0.47in}|p{0.64in}|p{0.97in}|p{1.49in}|}
\hline
\textbf{Component Benchmark} &  \textbf{Involved Motif} & \textbf{Application Domain} & \textbf{Workload Type} & \textbf{Data Set} & \textbf{Software Stack}\\
\hline

Xapian Server & Get, Put, Post & SE & Online service & Wikipedia entries & Xapian\\
\hline
PageRank & Matrix, Sort, Basic statistics, Graph & SE & Graph analytics & Google web graph & Hadoop, Spark, Flink, GraphLab, MPI\\
\hline
Index & Logic, Sort, Basic statistics, Set & SE & Offline analytics  & Wikipedia entries & Hadoop, Spark\\
\hline
Rolling top words & Sort, Basic statistics & SN & Streaming &  Random generate & Spark streaming, JStorm\\
\hline
\multirow{2}{*}{Kmeans} & \multirow{2}{*}{\tabincell{l}{Matrix, Sort, Basic statistics}} & \multirow{2}{*}{\tabincell{l}{SE, SN, EC,\\MP, BI}} & Offline analytics  & Facebook social network & Hadoop, Spark, Flink, MPI\\
\cline{4-6}
& & & Streaming  & Random generate & Spark streaming\\
\hline
\multirow{2}{*}{\tabincell{l}{Collaborative\\Filtering}} & \multirow{2}{*}{Graph, Matrix} & EC & Offline analytics  & Amazon movie review & Hadoop, Spark\\
\cline{3-6}
& & EC & Streaming & MovieLens dataset & JStorm\\
\hline
Naive Bayes & Basic statistics, Sort & SE, SN, EC & Offline analytics  & Amazon movie review & Hadoop, Spark, Flink, MPI\\
\hline
SIFT & Matrix, Sampling, Transform, Sort & MP & Offline analytics  & ImageNet & Hadoop, Spark, MPI\\
\hline
LDA & Matrix, Graph, Sampling & SE & Offline analytics  & Wikipedia entries & Hadoop, Spark, MPI\\
\hline
OrderBy & Set, Sort & EC & \multirow{2}{*}{\tabincell{l}{Data warehouse}}  & E-commerce transaction & Hive, Spark-SQL, Impala\\
\cline{1-3}\cline{5-6}
Aggregation & Set, Basic statistics & EC &  & E-commerce transaction & Hive, Spark-SQL, Impala\\
\hline
Project, Filter & Set & EC & \multirow{2}{*}{\tabincell{l}{Data warehouse}}  & E-commerce transaction & Hive, Spark-SQL, Impala\\
\cline{1-3}\cline{5-6}
Select, Union & Set & EC & & E-commerce transaction & Hive, Spark-SQL, Impala\\
\hline
Alexnet / Googlenet & \multirow{4}{*}{\tabincell{l}{Matrix, Transform,\\Sampling, Logic,\\Basic statistics}} & SN, MP, BI & AI  & Cifar, ImageNet & TensorFlow, Caffe, PyTorch\\
\cline{1-1}\cline{3-6}
Resnet / VGG16 & & SN, MP, BI & AI & Cifar, ImageNet & TensorFlow, Caffe, PyTorch\\
\cline{1-1}\cline{3-6}
Inception Resnet V2 & & SN, MP, BI & AI & Cifar, ImageNet & TensorFlow, Caffe, PyTorch\\
\cline{1-1}\cline{3-6}
DCGAN / WGAN & & SN, MP, BI & AI  & LSUN & TensorFlow, Caffe, PyTorch\\
\cline{1-1}\cline{2-6}
GAN & \multirow{2}{*}{\tabincell{l}{Matrix, Sampling, Logic,\\Basic statistics}} & SN, MP, BI & AI  & LSUN & TensorFlow, Caffe, PyTorch\\
\cline{1-1}\cline{3-6}
Seq2Seq & & SE, EC, BI & AI  &  TED Talks & TensorFlow, Caffe, PyTorch\\
\hline
Word2vec & Matrix, Basic statistics, Logic & SE, SN, EC & AI  & Wikipedia entries, Sogou data & TensorFlow, Caffe, PyTorch\\
\hline

\hline


\end{tabular}


\end{table*}

\subsubsection{Representative Real-world Data Set}
To cover a full spectrum of data characteristics, 
we collect 13 representative data sets, including different data sources (text, table, graph, and image), and data types of structured, un-structured, semi-structured.
Further, big data generation tools are provided to suit for different cluster scales, including text, table, matrix and graph generators.

\textbf{Wikipedia Entry} \cite{wikipedia} 
is a unstructured data set, consisting of 4,300,000 English articles.

\textbf{Amazon Movie Review} \cite{amazonreview} is
a semi-structured data set, consisting  of 7,911,684 reviews on 889,176 movies by 253,059 users. 

\textbf{Google Web Graph} (Directed  graph)\cite{googleweb} is a unstructured data set which
contains 875,713 nodes representing web pages and
5,105,039 edges representing the web links. 

\textbf{Facebook Social Graph} (Undirected graph) \cite{facebookgraph} 
contains 4,039 nodes, which represent users, and  88,234 edges, which
represent friendship between users. 

\textbf{E-commerce Transaction Data} is a structured data set from an e-commerce web site, consisting of two tables: ORDER
and ITEM. 


\textbf{\emph{ProfSearch} Person Resum\'{e}} is a semi-structured data set from a vertical search engine for scientists developed by ourselves, 
consisting of 278,956 resum\'{e}s automatically extracted from 20,000,000 web pages of about 200 universities and research institutions.

%

\textbf{CIFAR-10} \cite{krizhevsky2009learning} is a tiny image data set, which has 60,000 color images with the dimension of $32\times 32$. They are classified into 10 classes and each class has 6,000 examples.

\textbf{ImageNet} \cite{deng2009imagenet} is an image database organized according to the WordNet hierarchy.
We mainly use the Large Scale Visual Recognition Challenge 2014 (ILSVRC2014) \cite{russakovsky2014imagenet} data set. 

\textbf{LSUN} \cite{yu2015lsun} contains about one million labelled images, classified into 10 scene categories and 20 object categories.

\textbf{TED Talk} \cite{cettolo2012wit3} comes from translated TED talks, provided by IWSLT evaluation campaign.






\textbf{SoGou Data} \cite{sogouindex} is a unstructured data set, including corpus and search query data from Sogou Lab. The total data size is 4.98 GB. 

\textbf{MNIST} \cite{mnist} is a database of handwritten digits. It has a training set of 60,000 examples, and a test set of 10,000 examples. 

\textbf{MovieLens Dataset} \cite{harper2016movielens} is score data for movies, which has 9,518,231 training examples and 386,835 test examples (semi-structured text).

\subsubsection{State-of-the-art Techniques}

To perform apple-to-apple comparisons, 
we provide diverse implementations using the state-of-the-art techniques. 
For offline analytics, we provide Hadoop~\cite{hadoopweb}, Spark~\cite{zaharia2010spark}, Flink~\cite{mika2005flink} and MPI~\cite{mpich} implementations.
For graph analytics, we provide Hadoop, Spark GraphX~\cite{xin2013graphx}, Flink Gelly~\cite{flinkgelly} and GraphLab~\cite{low2012distributed} implementations.
For AI, we provide TensorFlow~\cite{abadi2016tensorflow}, Caffe~\cite{jia2014caffe} and PyTorch~\cite{pytorch} implementations.
For data warehouse, we provide Hive~\cite{thusoo2009hive}, Spark-SQL~\cite{sparksql} and Impala~\cite{bittorf2015impala} implementations.
For NoSQL, we provide MongoDB~\cite{chodorow2013mongodb} and HBase~\cite{george2011hbase} implementations.
For streaming, we provide Spark streaming~\cite{sparkstreaming} and JStorm~\cite{jstorm} implementations.

The Hadoop version of matrix multiplication benchmark is implemented based on Mahout~\cite{mahout}, and the Spark version is using Marlin~\cite{gu2015efficient}.
For AI, we identify representative and widely used data motifs in a wide variety of deep learning networks (i.e. convolution, relu, sigmoid, tanh, fully connected, max/avg pooling, cosine/batch normalization and dropout) and then implement each single motif and motif combinations as micro benchmarks and component benchmarks. The AI component benchmarks include Alexnet~\cite{krizhevsky2012imagenet}, Googlenet~\cite{szegedy2015going}, Resnet~\cite{he2016deep}, Inception\_Resnet V2~\cite{szegedy2017inception}, VGG16~\cite{simonyan2014very}, DCGAN~\cite{radford2015unsupervised}, WGAN~\cite{arjovsky2017wasserstein}, Seq2Seq~\cite{sutskever2014sequence} and Word2vec~\cite{mikolov2013distributed}, which are important state-of-the-art networks in AI.

\section{Experiment Setup}
In this section, we present our experiment configurations and methodology on characterizing the processor pipeline efficiency of big data and AI, in comparison to traditional benchmarks including SPEC CPU2006, PARSEC, and HPCC.

\subsection{Experiment Configurations}\label{config4}
We run a series of characterization experiments using \emph{BigDataBench 4.0} to obtain insights for architectural studies. From \emph{BigDataBench 4.0}, we test a majority of micro and component benchmarks with all seven workload types.

\begin{table}
\scriptsize
\caption{Configuration Details of Xeon E5-2620 V3}\label{hwconfigeration}
\renewcommand\arraystretch{1.2}
\center
\begin{tabular}{|p{0.66in}|p{0.66in}|p{0.66in}|p{0.66in}|}
\hline \rowcolor{mygray} \multicolumn{4}{|l|}{Hardware Configurations}\\
\hline \multicolumn{2}{|c|}{CPU Type} & \multicolumn{2}{c|}{Intel CPU Core} \\
\hline \multicolumn{2}{|c|}{Intel \textregistered Xeon E5-2620 V3}  &\multicolumn{2}{c|}{12 cores@2.40G} \\
\hline L1 DCache &L1 ICache &L2 Cache &L3 Cache \\
\hline 12 $\times$ 32 KB& 12 $\times$ 32 KB&12 $\times$ 256 KB& 15MB \\
\hline \multicolumn{2}{|c|}{Memory} & \multicolumn{2}{c|}{64GB,DDR4}  \\
\hline \multicolumn{2}{|c|}{Disk} & \multicolumn{2}{c|}{SATA@7200RPM}\\
\hline \multicolumn{2}{|c|}{Ethernet} & \multicolumn{2}{c|}{1Gb}\\
\hline \multicolumn{2}{|c|}{Hyper-Threading} & \multicolumn{2}{c|}{Disabled}\\
\hline
\end{tabular}
\end{table}

Our benchmarks support large-scale cluster deployments. For example, our industrial partner Huawei has evaluated the FusionInsight system on 12-node~\cite{fusion12} and 200-node~\cite{fusion200} clusters. In our experiments, we deploy an one-master-two-slave cluster for architecture evaluation, instead of a larger cluster because of the following reasons.
First, a larger cluster may lead to data skew which results in load unbalance in the cluster, and lead to the deviation of experimental results.
Second, the deployment and running cost is extremely high to collect all hardware events, which always need multiple times running to assure high accuracy of collected data for each benchmark~\cite{vtune}. A larger cluster aggravates the cost.
Third, most of previous architecture researches~\cite{yasin2014deep,ferdman2011clearing,jia2017understanding} also use a small-scale cluster.

In our experiments, each slave node has two Xeon E5-2620 V3 processors equipped with 64 GB memory and 6 TB disk. The detailed hardware configuration of each node is listed in Table \ref{hwconfigeration}. The software and compiler configurations are as follows: CentOS 7.2 with Linux kernel 4.1.13, JDK 1.8.0\_65, Hadoop 2.7.1, Apache Mahout 0.10.2, Hive 0.9.0, HBase 1.0.1, Scala 2.10.4, Spark 1.5.2, Python 2.7.5, TensorFlow 1.0, GCC 4.8.5. The level of optimization is “-O2”. 
With regard to the input data, we use 100 GB data for offline analytics, except that matrix multiplication uses 10000*10000 matrix data. 
Data warehouse uses 100 GB E-commerce transaction data.
Graph analytics uses $2^{26}$-vertex graph data.
For AI, we use CIFAR-10 data set and run 10 epoches for Alexnet, Googlenet and Inception\_Resent V2. For Resnet, we run 10000 training steps for each training step takes a short time. Word2vec  uses text8 wikipedia corpus.
We evaluate HBase with ten million records using NoSQL read and write benchmarks.
Online service processes million searching requests.
Spark streaming takes thousands of seconds streaming data as input and considers 10 seconds streaming data as a batch to process.

\subsection{Experiment Methodology}

We adopt a Top-Down methodology~\cite{yasin2014top} to evaluate the pipeline efficiency of big data and AI, which identifies the bottlenecks in a hierarchical manner. At the uppermost level,
it classifies an issued micro operation into four categories of retiring, bad speculation, frontend bound and backend bound. Totally, it has five levels, drilling down on the sub-tree of each category.
Modern processors provide hardware performance counters to support micro-architectural level profiling.
We use Perf~\cite{perf}, a Linux profiling tool, to collect the 
hardware events referring to the Intel Developer's Manual and pmu-tools~\cite{pmutools}.
To obtain more accurate performance counter values, we run each workload three times separately in order to sample the events during the whole runtime of workload. Then we report the average of the three runs.

\subsection{Compared Benchmarks Setup}
For comparison, we deploy SPEC CPU2006~\cite{spec2006spec}, PARSEC~\cite{bienia2008parsec} on one slave node and HPCC~\cite{luszczek2006hpc} on two slave nodes.

\textbf{SPEC CPU2006}: We run SPEC CPU 2006 with the reference input, reporting results averaged into two groups, i.e., integer benchmarks (\emph{SPECINT}) and floating point benchmarks (\emph{SPECFP}). The gcc version is 4.8.5.

\textbf{HPCC}: 
We run all seven HPCC benchmarks with version 1.4, including \emph{HPL}, \emph{STREAM}, \emph{PTRANS}, \emph{RandomAccess}, \emph{DGEMM}, \emph{FFT}, and \emph{COMM}.

\textbf{PARSEC}: We deploy PARSEC 3.0 Beta Release, which is a benchmark suite composed of multithreaded programs. We run all the 12 benchmarks with native input data sets and use gcc version 4.8.5 in compilation.

\section{Characterization Results}

We perform Top-Down analysis on seven types of big data and AI, drilling down on the five levels, and report our characterization results.
The seven types and corresponding software stacks include online service (Xapian), offline analytics (Hadoop, Spark), graph analytics (Hadoop, Spark), data warehouse (Hive, Spark sql), AI workloads (TensorFlow), NoSQL (HBase) and streaming (Spark streaming). For each software stack, we also report their average value of all benchmarks listed as AVG bar (e.g. Hadoop-AVG). In the rest of paper, we use Inception to represent Inception\_Resnet V2 benchmark. We run all workloads in traditional benchmarks, and present their average value, respectively. They are listed as SPECCPU-Int, SPECCPU-Float, PARSEC-AVG and HPCC-AVG, respectively.

\begin{figure}[tb]
\centering
\includegraphics[scale=0.35]{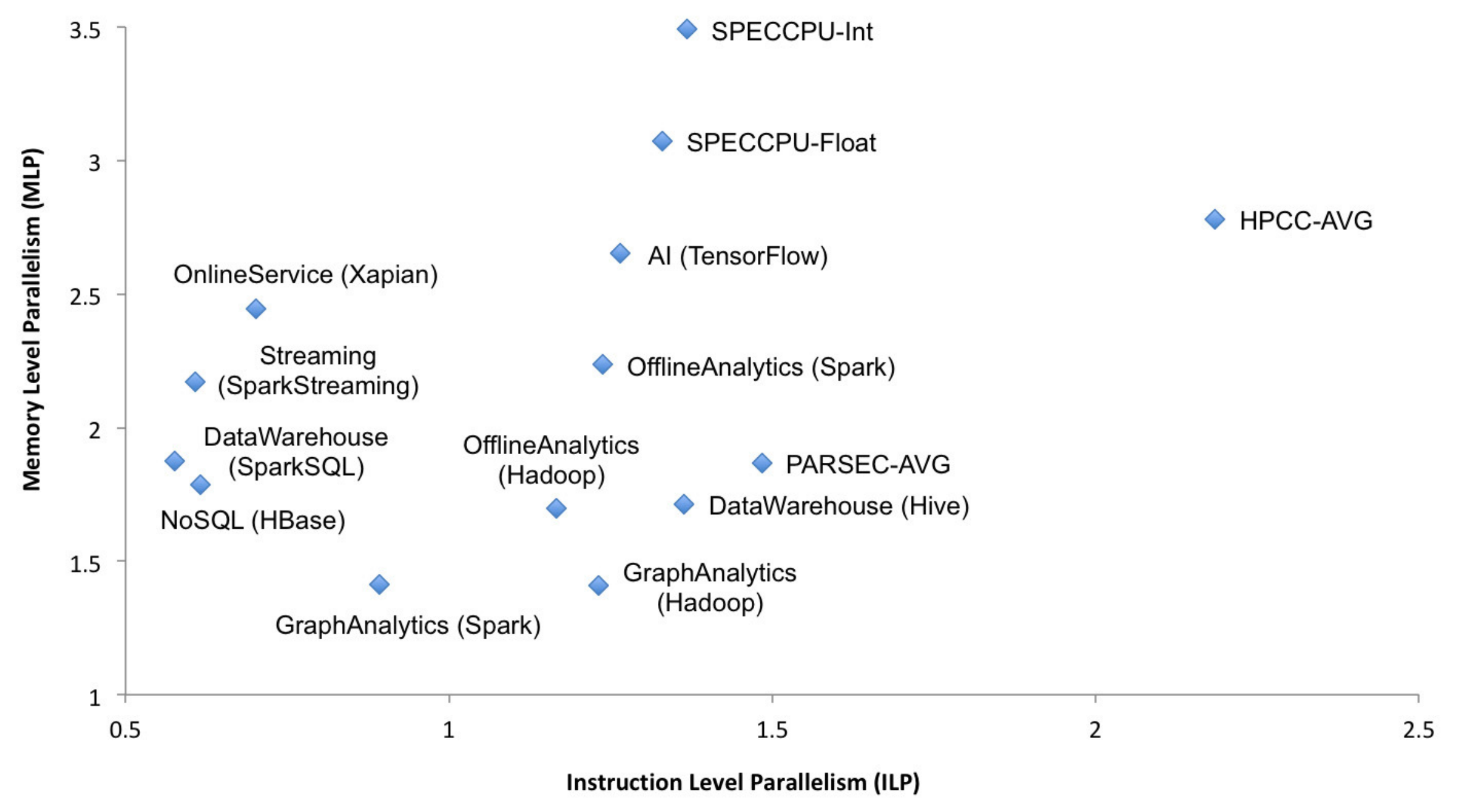}
\caption{Average Execution Performance.} 
\label{performance}
\end{figure}

In the rest of paper, we distinguish the software stacks for the same workload type when they have different behaviors, otherwise we only use the workload type to represent all software stacks when they reflect consistent behaviors.
The average execution performance of each workload type is shown in Fig. ~\ref{performance}, from the perspectives of ILP (instruction-level parallelism) and MLP (memory-level parallelism). We use IPC (retired instructions per cycle) to reflect the instruction level parallelism. 
MLP is measured as the average number of memory accesses when there is at least one such access, which indicates the dependencies of missing loads~\cite{chou2004microarchitecture}. As shown in Fig. ~\ref{performance},
different workload types or software stacks of big data reflect different execution performance. For example, the online service has low ILP while high MLP comparing to other types of big data, because it suffers from notable data cache misses and has low retired instruction percentage. 
Both the ILP and MLP of the AI are almost 1.5 times higher than that of big data on average. For several micro benchmarks of AI, such as Multiply and Pooling, their computations are simple and have little data dependencies, so they generate many concurrent data loads and incur many data cache misses~\cite{gao2018motif}, thus AI has higher MLP than big data. However, comparing to traditional benchmarks, the ILP of AI is lower on average. This is because that the AI framework implementation considers little instruction mix balance and memory access locality, while the traditional benchmarks like HPCC provides some computation-intensive kernels which are optimized to fully utilize the hardware resources.



\begin{figure*}[tb]
\centering
\includegraphics[scale=0.65]{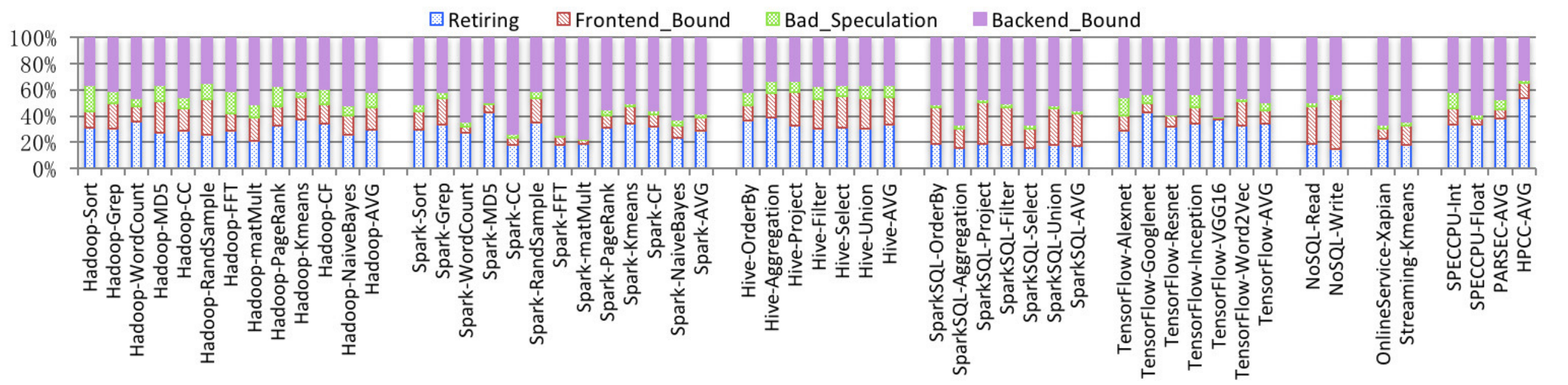}
\caption{Uppermost Level Breakdown of All Benchmarks.} 
\label{level1}
\end{figure*}


The uppermost-level breakdown of all benchmarks we evaluated are listed in Fig. ~\ref{level1}.
The retiring percentage of big data is 22.9\% on average, lower than traditional benchmarks (39.8\% on average),
which is also found by previous work~\cite{jia2017understanding} on Hadoop-based benchmarks. Specially, NoSQL, online service and streaming have extremely low retiring percentage, approximately 20\%. NoSQL has poor instruction locality, so it generates a large amount of instruction cache misses and greatly impact the performance. For online service and streaming, they suffer from notable backend stalls.
Further, we find that different workload types reflect diverse pipeline behaviors, indicating that they have different bottlenecks and need specific optimization strategies.
Corroborating the previous work~\cite{jia2017understanding}, backend bound is the first bottleneck and frontend bound is the second bottleneck for all big data we investigated. However, the frontend bound percentages vary across different workload types and software stacks. For example, eight out of twelve Spark-based benchmarks have low frontend bound percentages, only occupying less than 8\% on average.
NoSQL (about 35\%) and data warehouse (about 25\%) suffer from higher frontend bound than the others of big data (15\% on average) mainly because of instruction cache misses. 
In addition, software stacks and algorithms both have great impacts on pipeline behaviors. For example, the frontend bound and bad speculation is 17\% and 11\% for Hadoop based benchmarks on average, while 9\% and 3\% for Spark based. Also, for the same software stack, the frontend bound percentage is 20\% for Spark grep, while 6\% for Spark FFT.

AI has higher retiring percentage (35\% on average) than big data, approximately equal to the traditional benchmarks (39.8\%).
Backend bound is the first bottleneck for AI, however, frontend bound is not always the second bottleneck.
For example, the percentage of frontend bound and bad speculation for Alexnet is 11\% and 14\%, respectively.
On average, from the uppermost level breakdown, the percentages of frontend (both about 9\%) and backend bound (49.7\% v.s. 45.1\%) of AI are close to traditional benchmarks, while their bottlenecks at a deeper level are different. Neural network structures  have a great impact on pipeline behaviors. For example,
the percentage of frontend bound and bad speculation for VGG16 is about 1\%, respectively, while the percentage of frontend bound and bad speculation for Alexnet is more than 10\%, respectively. This is because that VGG16 have much more consecutive convolution computations than Alexnet.


Deeper analysis for each category is performed in the following subsections.

\subsection{Retiring}\label{retiring}

A pipeline slot represents hardware resources needed to process one micro operation~\cite{pipeslot}.
Retiring means pipeline slots fraction utilized by useful
work~\cite{yasin2014top}. Optimizing retiring percentage often increases the IPC metric and thus improves the execution efficiency. Retiring is composed of retiring regular uops and retiring uops fetched by the Microcode Sequencer (MS) unit.
MS unit is used to decode the CISC instructions which are not supported by the default decoders. However, the switches to MS unit have penalties and hurt performance~\cite{pmutools}.
We find that the 
numbers of uops decoded by MS unit of big data and AI are about 10 times larger than that of the traditional benchmarks. This result indicates that big data and AI have more CISC instructions needing microcode assists, which may suffer from more switch stalls and hurt performance.

\begin{figure*}[tb]
\centering
\includegraphics[scale=0.65]{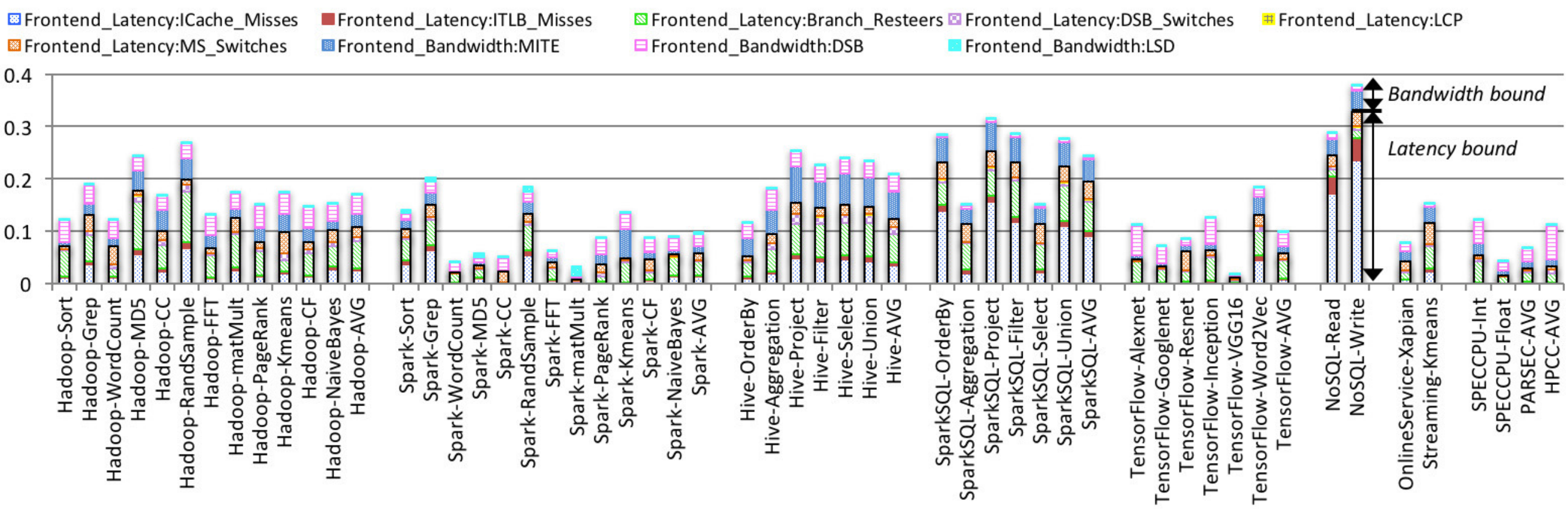}
\caption{Frontend Bound Breakdown (Level 2 \& 3) of All Benchmarks.} 
\label{l3frontend}
\end{figure*}

\subsection{Bad Speculation}

Bad speculation means slots fraction wasted due to incorrect speculations, including branch misprediction and machine clears. From our experimental results, we find that machine clears occupy about 0.1\% percentage for all benchmarks. Bad speculation mainly occurs due to branch misprediction, and their percentages are nearly equal to \emph{Bad\_Speculation} value in Fig. ~\ref{level1}.
Overall,  big data and AI have a small fraction of bad speculation, about 10\% for Hive and Hadoop benchmarks, 3\% for the other types and software stacks we evaluated.
For AI, different neural networks own different percentages of bad speculation, with 6\% on average.


\subsection{Frontend Bound}

Frontend bound occurs when frontend undersupplies the backend in a cycle. It is composed of two categories -- frontend latency bound (i.e. delivers no uops) and frontend bandwidth bound (i.e. delivers non-optimal amount of uops).
Fig. ~\ref{l3frontend} presents the frontend bound breakdown. Note that the y-axis of the black-bordered box indicates the percentage of frontend latency bound, and the length upper the black-bordered box indicates the percentage of frontend bandwidth bound. Taking Hadoop-Sort as an example, its frontend bound occupies a proportion of 12\%, with 7\% for latency bound and 5\% for bandwidth bound.
From Fig. ~\ref{l3frontend} we find that,
big data has more severe frontend bound than the traditional benchmarks, especially frontend latency bound, which is also found by previous work~\cite{ferdman2011clearing,kanev2015profiling,jia2017understanding}.
However, the frontend bound percentage varies across different workload types. Big data suffers from more frontend bound due to two reasons. First, the software stack changes the programming type comparing to original algorithm implementations, such as map/reduce interfaces in Hadoop. Second, the software stack itself incurs much more instructions, so the frontend bears the pressures of fetching and decoding these instructions.
AI benchmarks have different frontend bound percentages in terms of their layers and computation kernel proportions.
Frontend latency bound and bandwidth bound contribute to frontend bound equally.
Different from previous work~\cite{ferdman2011clearing,kanev2015profiling,jia2017understanding} that mainly identified frontend inefficiencies due to high instruction miss ratios and long latency introduced by caches, we thoroughly drill down on the sub-tree of frontend latency and bandwidth bound.


\subsubsection{Frontend Latency Bound}


Frontend latency bound indicates that frontend delivers no uops to backend, which may occur due to six reasons, including ICache misses, ITLB misses, branch resteers, DSB (decoded stream buffer) switches, LCP (length changing prefixes), and MS (microcode sequencer) switches. Among them, \emph{ICache misses} means stalls due to instruction cache misses. \emph{ITLB misses} means stalls due to instruction tlb misses. \emph{Branch resteers} means stalls due to frontend delay when fetching instruction from correct path, which may occur because of branch mispredictions. \emph{DSB switches} means stalls due to switches from DSB to MITE (Micro-instruction Translation Engine) pipelines. DSB is a decoded ICache used to store uops that have been decoded, so as to avoid penalties of legacy decode pipeline, which is also called MITE. \emph{DSB switches} are used to measure the penalties of switching from DSB to MITE~\cite{pmutools}. \emph{LCP} means stalls due to length changing prefixes, which can be avoided by using compiler flags. \emph{MS switches} means stalls due to switches of delivering uops to microcode sequencer. As mentioned in Subsection ~\ref{retiring}, retiring includes retiring regular uops and uops fetched by the MS unit. Generally, uops are coming from DSB or MITE pipeline. For some CISC instructions which cannot be decoded by default decoders, they must be handled by MS unit. However, frequent MS switches hurt performance, so \emph{MS switches} metric measures this penalties.

The breakdown within the black-bordered box in Fig. ~\ref{l3frontend} shows the proportions of the above six reasons that incur the frontend latency bound. We find that for big data except NoSQL, \emph{Branch resteers}, \emph{ICache misses} and \emph{MS switches} are three main reasons for frontend latency bound, while for NoSQL, the main reasons are \emph{ICache misses}, \emph{ITLB misses} and \emph{MS switches}.
The main reason of AI that incurs frontend latency bound is \emph{Branch resteers}, and the second reason is \emph{MS switches}, indicating that big data and AI indeed have much larger retiring uops from MS unit. 


\subsubsection{Frontend Bandwidth Bound}


\begin{figure*}[tb]
\centering
\includegraphics[scale=0.65]{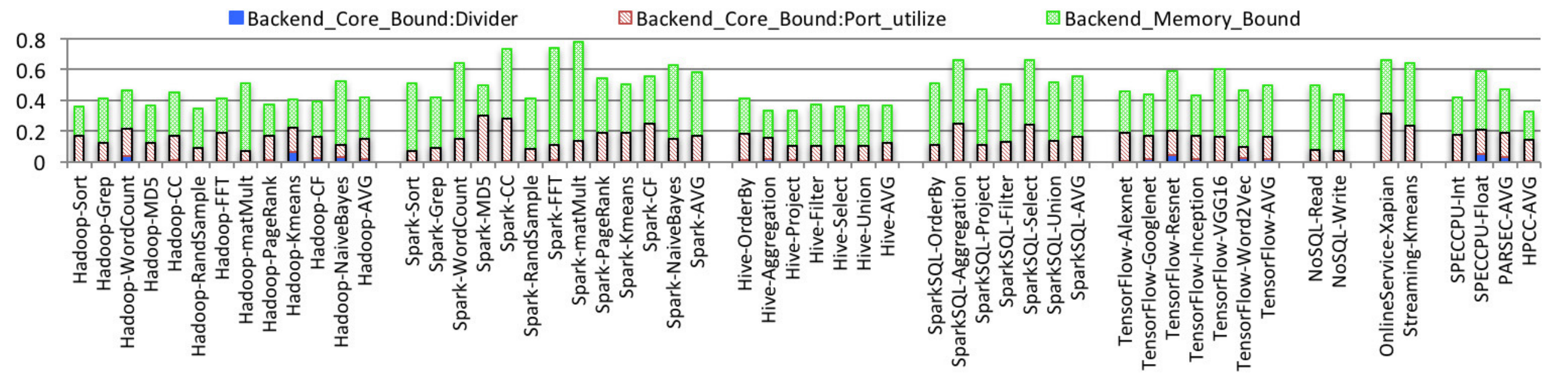}
\caption{Backend Bound Breakdown (Level 2) of All Benchmarks.} 
\label{l2backend}
\end{figure*}

\begin{figure*}[tb]
\centering
\includegraphics[scale=0.65]{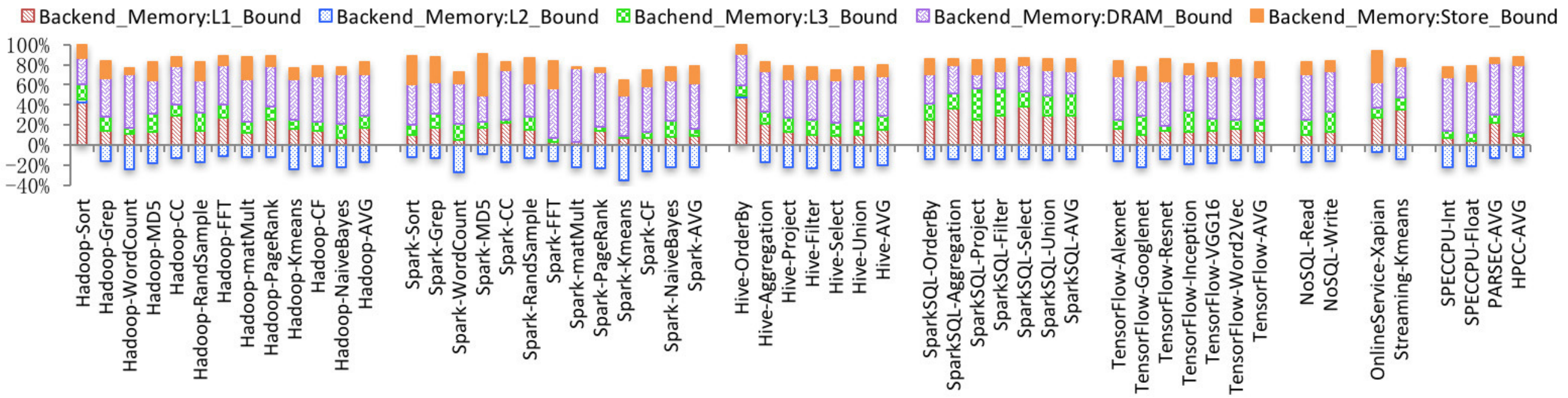}
\caption{Backend Memory Bound Breakdown (Level 3) of All Benchmarks.} 
\label{l3backmem}
\end{figure*}

Frontend bandwidth bound indicates the amount of uops delivering to backend is less than theoretical value, such as four for Haswell architecture. The frontend bandwidth bound is mainly due to three reasons, including \emph{MITE}, \emph{DSB} and \emph{LSD}. Among them, \emph{MITE} means stalls due to \emph{MITE} pipeline, such as the inefficiency of the instruction decoders. \emph{DSB} means stalls due to \emph{DSB} fetch pipeline, such as inefficient utilization of DSB. \emph{LSD} means stalls due to loop stream detector unit, which occupies a little generally.

The breakdown of frontend bandwidth bound in Fig. ~\ref{l3frontend} shows the proportions of the above three reasons. \emph{DSB} and \emph{MITE} are two main reasons for nearly all listed benchmarks.
However, different workload types have different first frontend bandwidth bottleneck.
For offline analytics and graph analytics, their first frontend bandwidth bottleneck is \emph{DSB}. For data warehouse, NoSQL, online service and streaming, their first frontend bandwidth bottleneck is \emph{MITE}.
For AI, their first bottleneck of frontend bandwidth bound is \emph{DSB}, except \emph{MITE} for Word2Vec benchmark.
In order to reduce the frontend bandwidth bound and improve the performance of big data and AI, DSB utilization and MITE pipeline efficiency need to be optimized.


\subsection{Backend Bound}

Backend bound occurs when the backend has not enough required resources to process new uops, which can be divided into backend core bound and backend memory bound. Among them, backend core bound refers to non-memory core issues, such as the lack of out-of-order resources. Backend memory bound means the stalls due to load or store instructions.

Fig. ~\ref{l2backend} lists the backend bound breakdown of all benchmarks. 
The black-bordered boxes indicate the percentage of backend core bound slots, and the green boxes above them indicate the percentage of backend memory bound slots.
The first bottleneck of big data and AI is backend bound. Previous work~\cite{jia2017understanding} found core bound and memory bound nearly contribute to the backend bound equally. However, we find memory bound is more severe than core bound for all big data and AI benchmarks, except that online service has nearly equal core bound and memory bound.


\begin{figure*}[tb]
\centering
\includegraphics[scale=0.65]{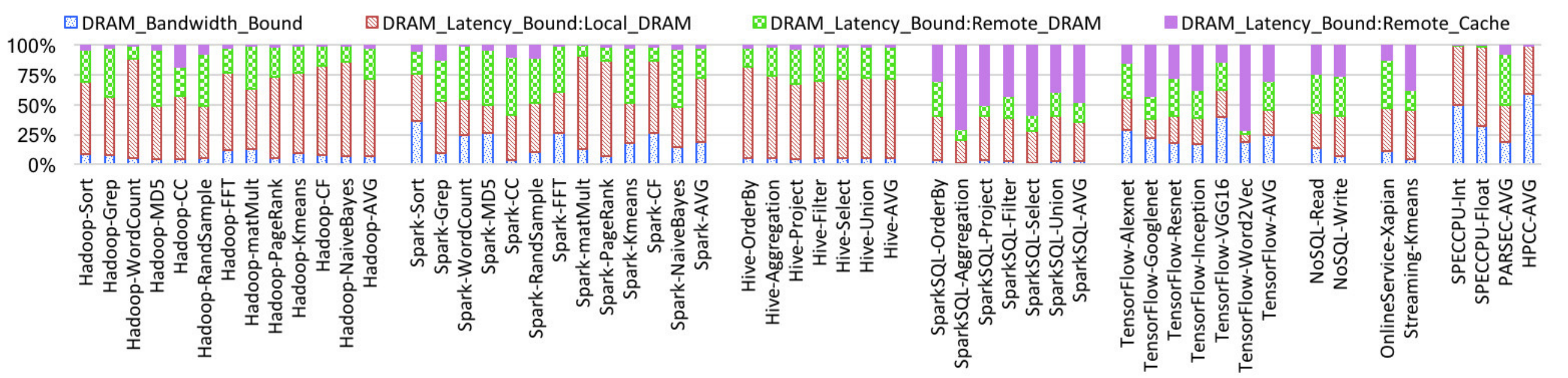}
\caption{DRAM Bound Breakdown (Level 4) of All Benchmarks.} 
\label{l4dram}
\end{figure*}


\subsubsection{Backend Core Bound}

Backend core bound can further be split into \emph{Divider} and \emph{Port utilization}. \emph{Divider} means the cycle fraction that the Divider unit is in use, which has longer latency than other integer or floating-point operations. \emph{Port utilization} means the stalls due to low utilization of execution ports. For example, Haswell has eight execution ports, and each port can execute specific uops (four ports for computation and four ports for load/store operations).
These execution ports may be under-utilized in a cycle due to data dependency of instructions or non divider-related resource contention~\cite{pmutools}.

The breakdown within the black-bordered box in Fig. ~\ref{l2backend} shows proportions of \emph{Divider} and \emph{Port utilization} that incur the backend core bound. \emph{Divider} occupies a small proportion, except for some computation intensive workloads, such as Hadoop Kmeans. From Fig. ~\ref{l2backend} we find the utilizations of execution ports are low for big data and AI, further indicating that the instruction mix balance need to be improved.

\subsubsection{Backend Memory Bound}

Backend memory bound can further be divided into \emph{L1 Bound}, \emph{L2 Bound}, \emph{L3 Bound}, \emph{DRAM Bound}, and \emph{Store Bound}, which incurs stalls related to memory hierarchy.

Fig. ~\ref{l3backmem} shows the normalized backend memory bound breakdown. Note that \emph{L2 Bound} is negative due to PMU erratum on L1 prefetchers~\cite{yasin2014top}.
We find that the main reason for backend memory bound is \emph{DRAM Bound} for big data and AI, except that online service suffers from more store bound than DRAM bound.
Different from the traditional benchmarks, big data and AI also suffer from more stalls due to \emph{L1 Bound}, \emph{L3 Bound} and \emph{Store Bound}.


\emph{DRAM Bound} is the first Backend Memory Bound bottleneck for most benchmarks, and we further analyze two factors that incur \emph{DRAM Bound}, including DRAM latency and DRAM bandwidth. DRAM latency means the stalls due to the latency from dynamic random access memory, it can be further classified into stalls due to loads from local memory (\emph{Local\_DRAM)}, remote memory (\emph{Remote\_DRAM}) and remote cache (\emph{Remote\_Cache}). DRAM bandwidth means the stalls due to memory bandwidth limitations.
Fig. ~\ref{l4dram} presents the DRAM bound breakdown, including DRAM bandwidth bound and three kinds of DRAM latency bound---local DRAM, remote DRAM latency and remote cache. 
Different from the traditional benchmarks, the first DRAM bound bottleneck of big data and AI is DRAM latency bound. 
AI suffers from more DRAM bandwidth bound than big data.
In terms of DRAM latency bound, the main reason for big data is local DRAM latency on average, except that Spark sql suffers from more remote cache latency. Also, the main reason for AI is remote cache latency. Remote cache or remote DRAM latencies are mainly due to non-optimal NUMA allocations. Processor affinity and NUMA-friendly data placement may reduce the latency and improve the performance.

\subsection{Discussion on AI Benchmarks}

To explore the performance of AI benchmarks considering different hardware architectures and running configurations, we first characterize them on GPUs and then evaluate the impacts of iteration numbers on architecture behaviors.

\subsubsection{AI Benchmarks on GPUs}

\begin{figure*}[tb]
\centering
\includegraphics[scale=0.45]{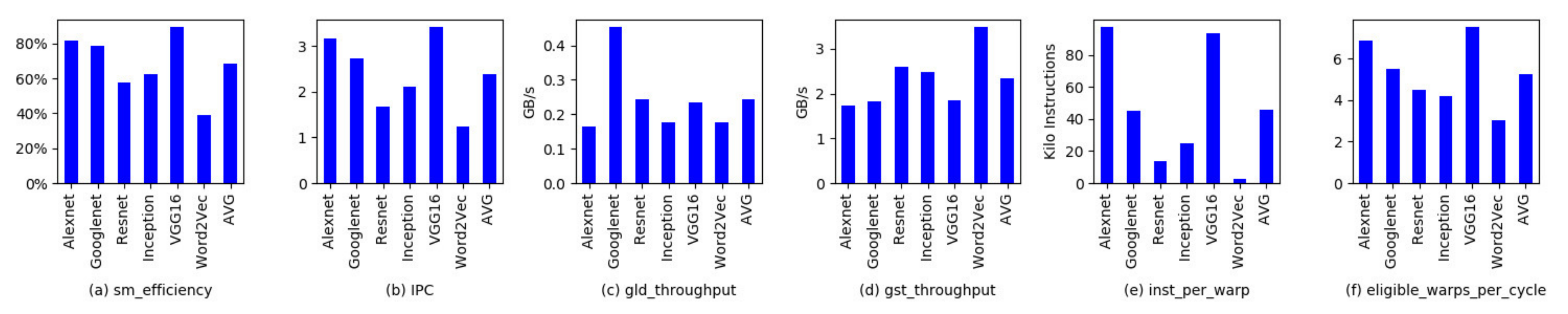}
\caption{Performance of AI Benchmarks on GPUs.} 
\label{gpu}
\end{figure*}

Since a significant portion of the computationally intensive AI tasks are performed on GPUs, we further evaluate AI benchmarks on GPUs, using PAPI CUDA Component~\cite{papi} and CUDA profiling tool---nvprof~\cite{nvprof}. The CPU platform is the same with Section~\ref{config4}. The GPU platform is NVIDIA Tesla K80 with two Tesla GK210 GPUs. Each GPU has 13 stream multiprocessors (SM), and each SM includes 192 cores. The total memory is 24 GB GDDR5.

SM efficiency and IPC are two important metrics to evaluate the execution performance on GPUs. Among them, SM efficiency indicates the percentage of time that the SM has one or more warps are active. IPC means the instructions executed per cycle. We evaluate the AI benchmarks on GPUs, as shown in Fig.~\ref{gpu}(a) and (b), different neural network structures reflect different performance on GPUs. Resnet and Word2Vec have lower IPC and SM efficiency than other AI benchmarks. We further evaluate their memory access and computation patterns to explore why they reflect different performance. Fig.~\ref{gpu}(c) and (d) show their global memory load and store throughput, respectively. Fig.~\ref{gpu}(e) and (f) present their average number of instructions executed by each warp and the average number of warps that are eligible to be issued per cycle, respectively. We find that Resnet, Inception and Word2Vec have higher memory access requirements, so they suffer from more stalls due to data load and store, and thus they have insufficient instructions or warps to be executed.


\begin{figure}[tb]
\centering
\includegraphics[scale=0.55]{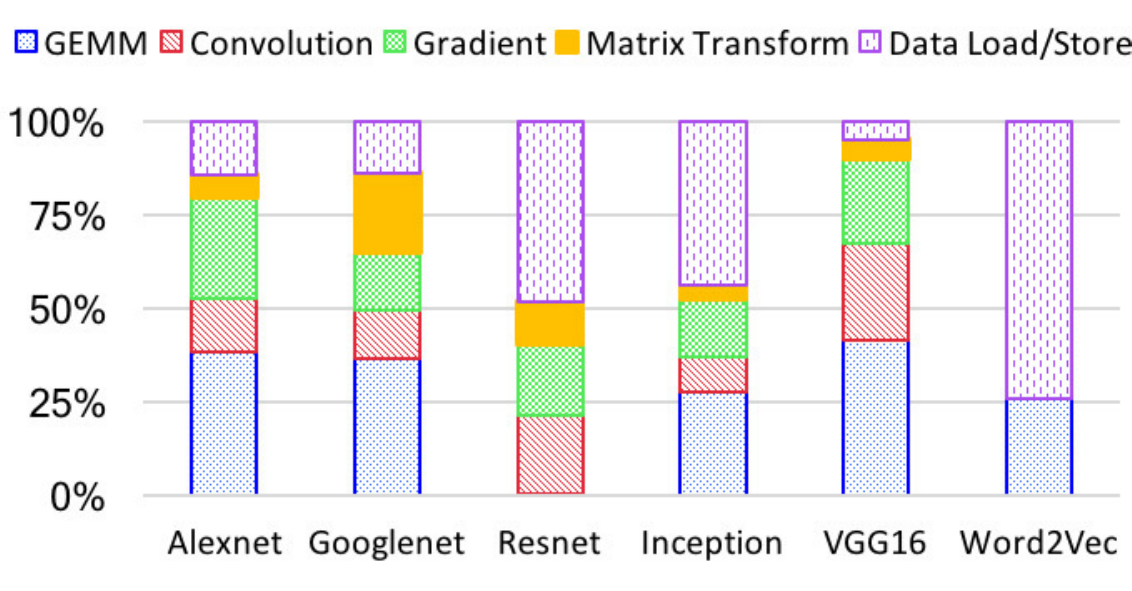}
\caption{Runtime Breakdown of Kernels.} 
\label{gpukernel}
\end{figure}

To explore the reasons why they have different computation and memory access patterns, we analyze the runtime breakdown of kernels within each AI benchmark.
As shown in Fig.~\ref{gpukernel}, we classify the most time-consuming kernels into five categories---GEMM, Convolution, Gradient, Matrix Transform, and Data load/store. Among them, GEMM represents the matrix multiplication kernels, including cgemm kernels and sgemm kernels. Convolution means the convolution-related kernels, including convolve, fft and winograd kernels. Gradient indicates the gradient computations, including dgrad engine and wgrad engine. For example, the backward kernel of convolution belongs to this category. Matrix transform includes matrix transpose, pooling and normalization. Data load/store includes the kernels that perform data load and store operations, such as memcpy and tensor evaluator. From Fig.~\ref{gpukernel}, we find Alexnet, Googlenet and VGG16 spend a majority of runtime on computation kernels and involve in little data loads and stores, so they have the highest IPC and SM efficiency. 
However, Resnet, Inception and Word2Vec spend too much time on data movements.
Resnet and Inception use batch normalizations to speed up deep neutral networks, while Alexnet and Googlenet use quite a few local response normalizations (LRN).
Batch normalization calls \emph{assign\_moving\_avg} to update variables, which has many data loads and stores, much larger than that of LRN. So Resnet and Inception spend much time on data movements and further impact performance. In addition, Resnet has deeper layers and mainly uses winograd algorithm to compute convolution, while the others either use fft and matrix multiplication to compute convolution or have more simple structure, so Resnet spends less time on GEMM kernels than others.

In conclusion, the memory access patterns impact the performance on GPUs greatly. The performance can be improved from two levels. From the GPU architecture design level, the stalls due to memory accesses need to be optimized. From the application level, the implementation of kernels and frameworks need to consider more efficient memory access, such as better locality.

\subsubsection{Iteration Impact on Architecture Behaviors}

\begin{figure}[tb]
\centering
\includegraphics[height=1.8in, width=3in]{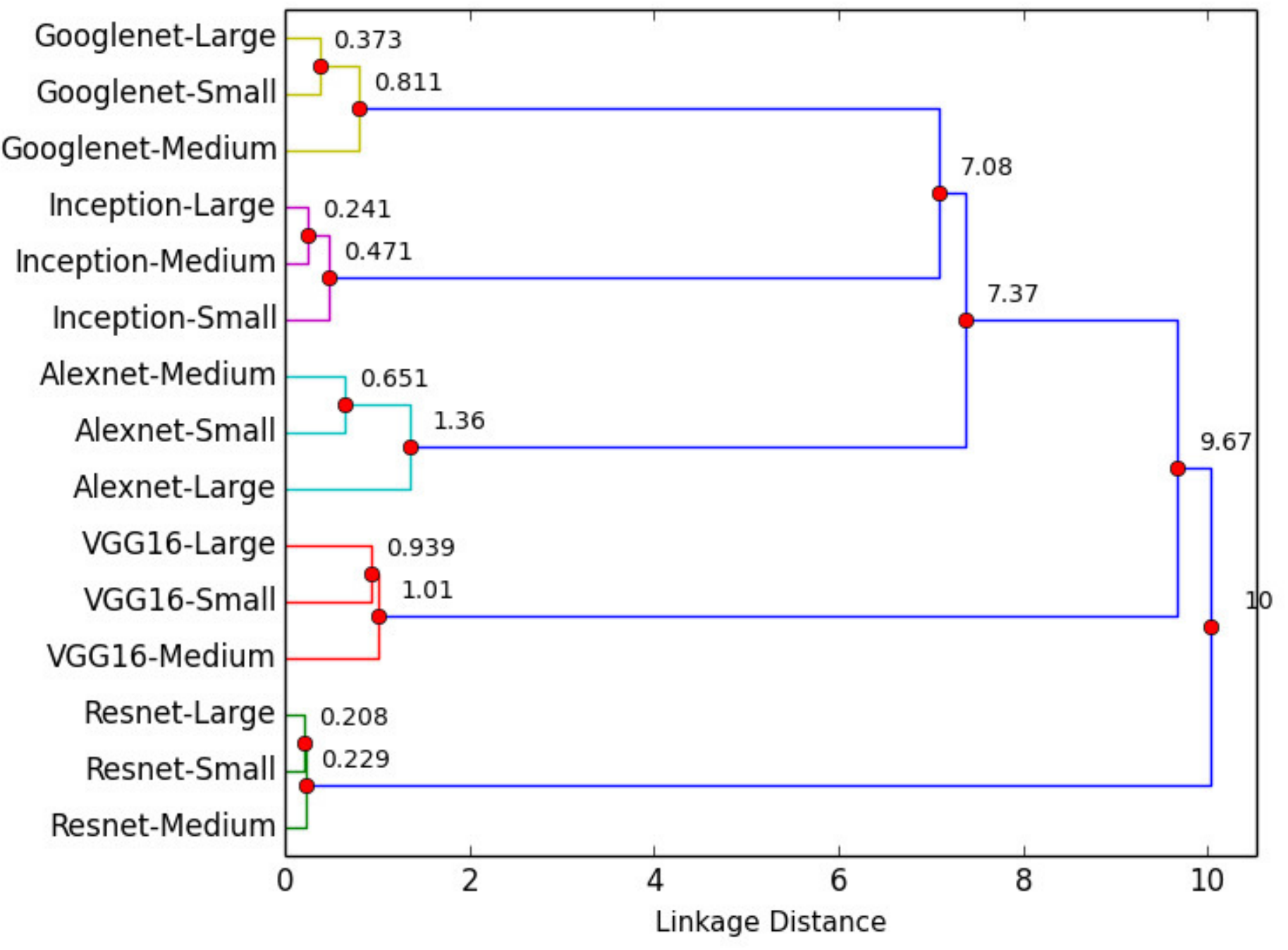}
\caption{Similarity with Different Iterations.} 
\label{iterimpact}
\end{figure}


AI benchmarks always need hundreds of iterations to obtain higher prediction precision and lower training loss. However, for architecture research, AI benchmarks are too time-consuming even if running on GPUs. To evaluate the impact of iteration number on microarchitectural characteristic  of  AI, we run five neural networks using different number of iterations -- \emph{Small, Medium, Large}. For Alexnet, Googlenet, Inception and VGG16 networks, we run 1 (\emph{Small}), 10 (\emph{Medium}), 100 (\emph{Large}) epoches, respectively. For Resnet networks, we run 2000 (\emph{Small}), 10000 (\emph{Medium}), 50000 (\emph{Large}) training steps. respectively.
We use PCA~\cite{jolliffe1986principal} and hierarchical clustering~\cite{johnson1967hierarchical} to measure the similarity, using all fifty micro-architectural metrics we collect according to the Top-Down method.
Fig. ~\ref{iterimpact} presents the linkage distance of all AI benchmarks, and the smaller distance means the higher similarity.
We find that the same neural networks with different iteration numbers are clustered together and have shorter distance, which means a small number of iterations is enough for micro-architectural evaluation of AI benchmarks.

\section{Conclusion}

In this paper, we propose a data motif-based scalable benchmarking methodology to build micro, component, and end-to-end application benchmarks.
Following this methodology, we set up a unified open source big data and AI benchmark suite -- BigDataBench 4.0.
Finally, we comprehensively characterize BigDataBench 4.0 on CPUs and GPUs, respectively.

\ifCLASSOPTIONcaptionsoff
  \newpage
\fi

\end{document}